\documentclass[aps,onecolumn,qic,superscriptaddress]{revtex4-1}
\usepackage[unicode=true,pdfusetitle, bookmarks=true,bookmarksnumbered=false,bookmarksopen=false, breaklinks=false,pdfborder={0 0 0},backref=false,colorlinks=false] {hyperref}
\hypersetup{colorlinks,linkcolor=myurlcolor,citecolor=myurlcolor,urlcolor=myurlcolor}
\usepackage{graphics,epstopdf,graphicx,amsthm,amsmath,amssymb,times,braket,color,bm}
\usepackage[up]{subfigure}
\usepackage{cleveref}
\usepackage{mathrsfs}

\definecolor{myurlcolor}{rgb}{0,0,0.7}
\newcommand{\proj}[1]{| #1\rangle\!\langle #1 |}
\newcommand{\setft}[1]{\mathrm{#1}}
\newcommand{\lin}[1]{\setft{L}\left(#1\right)}
\newcommand{\density}[1]{\setft{D}\left(#1\right)}
\newcommand{\Br}[1]{\left[#1\right]}

\newcommand{\norm}[1]{\left\lVert #1 \right\rVert}
\newcommand{\abs}[1]{\left\lvert #1 \right\rvert}
\newcommand{\Pa}[1]{\left(#1\right)}

\def \diag {\mathrm{diag}}
\def\ot{\otimes}
\def\complex{\mathbb{C}}

\def\cC{\mathcal{C}}
\def\cH{\mathcal{H}}\def\cI{\mathcal{I}}
\def\cK{\mathcal{K}}

\def\rD{\mathrm{D}}

\def\rS{\mathrm{S}}

\def\sC{\mathscr{C}}

\theoremstyle{plain}
\newtheorem{thm}{\protect\theoremname}
\newtheorem{prop}[thm]{Proposition}
\newtheorem{remark}[thm]{Remark}
\newtheorem{lem}[thm]{Lemma}
\providecommand{\theoremname}{Theorem}
\newcommand*{\myproofname}{Proof}

\makeatother

\begin{document}


\title{Cohering power of quantum operations}


 \author{Kaifeng Bu}
 \email{bkf@zju.edu.cn}
 \affiliation{School of Mathematical Sciences, Zhejiang University, Hangzhou 310027, PR~China}

 \author{Asutosh Kumar}
 \email{asukumar@hri.res.in}
 \affiliation{Harish-Chandra Research Institute, Chhatnag Road, Jhunsi, Allahabad 211019, India}
 \affiliation{Homi Bhaba National Institute,  Anushaktinagar, Mumbai 400094, India}

 \author{Lin Zhang}
 \email{linyz@zju.edu.cn}
 \affiliation{Institute of Mathematics, Hangzhou Dianzi University, Hangzhou 310018, PR~China}

 \author{Junde Wu}
 \email{wjd@zju.edu.cn}
 \affiliation{School of Mathematical Sciences, Zhejiang University, Hangzhou 310027, PR~China}


\begin{abstract}
Quantum coherence is a basic feature of quantum physics. Combined with tensor product structure of state space, it gives rise to the novel concepts such as entanglement and quantum correlations, which play a crucial role in quantum information processing tasks. However, quantum correlations, especially entanglement, are fragile under decoherence. In this context, very few investigations have touched on the production of quantum coherence by quantum operations. In this paper, we study cohering power -- the ability of quantum operations to produce coherence.
First, we provide an operational interpretation of cohering power. Then, we decompose a generic quantum operation into three basic operations, namely, unitary, appending and dismissal operations, and show that 
the cohering power of any quantum operation is upper bounded by the corresponding unitary operation.
Furthermore, we compare cohering power and generalized cohering power of quantum operations for different measures of coherence.
\end{abstract}

\maketitle

\section{Introduction}
Quantum superposition, arising from the linearity of quantum mechanics, is the most fundamental feature of quantum mechanics. It is one of the characteristic distinguishing properties between classical and quantum systems. It is responsible for almost all the intriguing quantum phenomena such as interference of microscopic particles. Quantum coherence \cite{Baumgratz2014}, which is identified by the presence of off-diagonal terms in the quantum states, is a direct consequence of the superposition principle. It builds the foundation of quantum theory. Moreover, combined
with the tensor product structure of quantum state space, quantum superposition can give rise to various quantum correlations including quantum entanglement \cite{HorodeckiRMP09}, which form an important physical resource in quantum information processing tasks \cite{Nielsen10}. Entangled states have vast applications in as many fields as quantum communication and computation, in quantum metrology \cite{Paris2009}. However, unlike entanglement and quantum correlations, quantum coherence is a basis-dependent quantity. That is, coherence of a given quantum state can be quite different within different reference frameworks. For example, while the state
$|\psi\rangle = \frac{1}{\sqrt{2}}(|0\rangle + |1\rangle) = |+\rangle$ has coherence in $\sigma_z$-basis
$\{|0\rangle,|1\rangle \}$, it has zero coherence in $\sigma_x$-basis
$\{|\pm\rangle = \frac{|0\rangle \pm |1\rangle}{\sqrt{2}}\}$. On the other hand, $|\phi^+\rangle = \frac{1}{\sqrt{2}}(|00\rangle + |11\rangle) = \frac{1}{\sqrt{2}}(|++\rangle + |--\rangle)$, is both coherent and entangled in both the above bases. The dependence of quantum coherence on the choice of basis--the reference basis--can sound disturbing at first, but it is naturally determined by the experimental situation at hand.
Quantum coherence has been widely applied in fields like quantum thermodynamics
\cite{AspuruG13,RudolphNat,Lostaglio2015,Brandao2015,Narasimhachar2015} and quantum biology
\cite{Plenio2008, Lloyd2011, Levi14}. These suggest coherence to be a useful resource at the nanoscale. It has also application in quantum parallelism \cite{Nielsen10}. These important advancements in quantum domain suggest that there should be a quantitative framework for coherence. Like entanglement, a rigorous framework for the quantification of quantum coherence, from a resource-theoretic point of view, has been developed recently in Ref. \cite{Baumgratz2014}. In any resource theory, there are two basic components: free (allowed) states and free (allowed) operations. The resource theory of quantum coherence, likewise, is based on the set of ``incoherent operations'' as the free operations and the set of ``incoherent states'' as the set of free states. As remarked earlier, the set of incoherent
states and the set of incoherent operations depend critically on the choice of basis.
Recently, a significant effort has been devoted towards quantifying quantum superposition, and hence quantum coherence, from a resource theoretic perspective \cite{Baumgratz2014,Aberg2006,Girolami14,Bromley2015,Alex15,Xi2015,Winter2015,Shao2015,Pires2015,Guo2015,
Yao2015,Killoran2015,Wu2015,Uttam2015,UttamA2015,Cheng2015,Mondal2015,Mondal2016,Kumar2015,Mani2015,
ChitambarPRL2016,Du15,StreltsovA2015,Chitambar2015,Bera2015,Bagan2015,Hillery2016,Rana2016,Rastegin2016}.
In order to exploit quantum coherence, we need to quantify coherence in a given state.
Along this line, several kinds of coherence measures such as
$l_1$-norm of coherence, relative entropy of coherence and skew-information of coherence have been introduced in \cite{Baumgratz2014,Girolami14}. That quantum coherence can be measured with entanglement was shown in Ref. \cite{Alex15}.
It turns out that a proper measure of coherence should satisfy following properties: (i) (\emph{Nullity}) Incoherent states have zero coherence, (ii) (\emph{Monotonicity}) Incoherent completely positive and trace preserving maps cannot increase coherence, and/or the average coherence should not increase under selective measurements, and (iii) (\emph{Convexity}) Non-increasing of coherence under the mixing of quantum states.
Besides, the relationship of coherence with other
quantities like mixedness and quantum discord were revealed in \cite{Uttam2015,Xi2015,Yao2015,Cheng2015}.
Several statistical properties of quantum coherence have
also been obtained in Refs. \cite{UttamS2015,ZhangLin2015} in parallel to entanglement.
Furthermore, the operational resource theory of quantum coherence has
been developed in \cite{Winter2015}, and the transformation processes
like coherence distillation have been studied to give physical
interpretations to the coherence measures mentioned above.

Quantum systems are notoriously different from classical systems,
and can outperform classical systems in many information-processing tasks,
including quantum communication and computation \cite{Nielsen10}.
Therefore, quantum technology has significant importance to information technology.
However, one of the major bottlenecks of quantum technology is the quantum decoherence effect \cite{Nielsen10}.
In the phenomenon of decoherence, a quantum system inevitably interacts with its surroundings and loses quantum coherence. Quantum correlations, especially entanglement, have been found to be fragile under decoherence \cite{Rivas2012,Rivas2014}. The decoherence effect inherently leads to the dissipation from quantum systems to classical systems. However, we need to avoid the phenomenon of decoherence when we
implement quantum techniques in quantum information and computation
\cite{Nielsen10}. Due to immense importance of quantum correlations and entanglement--being indispensable resources--in quantum information processing tasks, maintaining coherence and/or quantum correlations in quantum systems is a challenging assignment. Here, the cohering power of quantum operations turns up as a savior.
The cohering power of a quantum operation quantifies the ability to produce coherence. Authors in \cite{Mani2015} have calculated cohering power of some special qubit operations exactly.
In our work, we investigate two different types of cohering power of generic quantum operations.
First, we give an operational interpretation of cohering power.
Moreover, we address the problem of estimating cohering power of a generic quantum operation in terms of simple operations. We calculate the cohering power of
three basic quantum operations, namely, unitary operation, appending operation and dismissal operation.
Then, by dividing a generic quantum operation into these three quantum operations, we
obtain an upper bound on the cohering power of this generic operation in terms of
cohering power of these basic operations.
More importantly, we compare two different kinds of cohering powers ($\sC_{\cC}$ and
$\widehat{\sC}_{\cC}$) for unitary operations. For $l_1$-norm
measure, they coincide in single qubit case only, while they are
different for any number of qubits for relative entropy of coherence.

The paper is organized as follows. Introductory material about
coherence is presented in Sect.~\ref{sec.intro}. We give an information-theoretic
interpretation of cohering power in Sec.~\ref{sec:phy_mean}. In
Sec.~\ref{sec:cohe_pow}, we calculate cohering power of
three basic quantum operations, and obtain an upper bound on the cohering
power of a generic quantum operation in terms of that of these basic
quantum operations. Sec.~\ref{sec:diff_def} is devoted to comparison of
cohering powers, $\sC_{\cC}$ and
$\widehat{\sC}_{\cC}$, for different measures of coherence. Finally,
we conclude in Sec.~\ref{sec:conclu}.

\section{Preliminary and notations}\label{sec.intro} 

Throughout this paper, we assume that all quantum systems are
described by finite-dimensional complex Hilbert spaces. A quantum state
$\rho$ on $\cH$ is a positive semi-definite operator
with unit trace. The set of all quantum states on $\cH$ is
denoted by $\density{\cH}$. Besides, the operator $\rho_{\psi} =
\proj{\psi}$ is said to be a pure state for each unit vector
$\ket{\psi} \in \cH$.
Let $\cH$ be a $d$-dimensional complex Hilbert space and
$\{|k\rangle\}^d_{k=1}$ be some reference basis of $\cH$. For each quantum
state $\rho\in \density{\cH}$, identified as a matrix
$(\rho_{ij})$ with respect to the basis $\{|k\rangle\}$, we
have following two proper measures of coherence \cite{Baumgratz2014}:

\begin{enumerate}
\item[(i)] $ l_{1}$-norm of coherence, $\cC_{l_{1}}$, is defined by
\begin{eqnarray}
\cC_{l_{1}}(\rho):=\sum_{i\neq j}|\rho_{ij}|.
\end{eqnarray}
\item[(ii)] Relative entropy of coherence, $\cC_{r}$, is defined by
\begin{eqnarray}
\cC_{r}(\rho):= \mathrm{min}_{\sigma \in {\cal I}} \rS(\rho \parallel \sigma) = \rS(\rho_{\diag}) - \rS(\rho),
\end{eqnarray}
where ${\cal I}$ is the set of all incoherent states on $\cH$, that is, all states on $\cH$ which are diagonal matrices with respect to the reference basis $\{|k\rangle\}$, $\rS(\rho \parallel \sigma) = \mathrm{Tr}\rho(\log \rho - \log \sigma)$ is the relative entropy between $\rho$ and $\sigma$, $\rS(\rho)=-\mathrm{Tr}{\rho\log\rho}$ is the von Neumann entropy of
$\rho$, and $\rho_{\diag}$ is the quantum state obtained from $\rho$ by
deleting all the off-diagonal elements of $\rho$, that is, $\rho_{\diag} = \sum_k \langle k|\rho|k\rangle \proj{k}$.
\end{enumerate}

It follows from the definition of coherence measures that:
\begin{eqnarray}\label{multi}
\cC_{l_{1}}(\rho \ot \sigma)+1=
(\cC_{l_{1}}(\rho)+1)(\cC_{l_{1}}(\sigma)+1),
\end{eqnarray} and
\begin{eqnarray}
\cC_{r}(\rho \ot \sigma)=\cC_{r}(\rho)+\cC_{r}(\sigma).
\end{eqnarray}

Let $\lin{\cH, \cK}$ be the set of all linear operators from $\cH$
to $\cK$. If $\cH = \cK$, we denote $\lin{\cH, \cH}$ by
$\lin{\cH}$. A linear transformation $\Phi$ which maps $\lin{\cH}$ to
$\lin{\cK}$ is said to be a \emph{quantum operation} if there are
finite linear operators $K_{\mu}\subseteq \lin{\cH, \cK}$ such that
$\sum_{\mu} K^{\dag}_{\mu}K_{\mu}=\mathbb{I}$ and for each $\rho\in \density{\cH}$,
$$\Phi(\rho)=\sum_{\mu}
K_{\mu}\rho K^{\dag}_{\mu}.$$

The quantum operation $\Phi$ is said to be \emph{incoherent
operation} if for each
$K_{\mu}$, $K_{\mu}\cI K^{\dag}_{\mu}\subset \cI$ (upto a normalization) \cite{Baumgratz2014}.

Let $\cC$ denote the coherence measure $\cC_{l_1}$ or $\cC_{r}$.
Recall that the cohering power and the generalized cohering power of a quantum operation are defined respectively
by \cite{Yao2015}:
\begin{eqnarray}\label{cp}
\label{cp1}\sC_{\cC}(\Phi):&=&\max\set{\cC(\Phi(\delta)):\delta\in \cI}\nonumber\\
\label{cp3}&=&\max\set{\cC(\Phi(\delta)):\delta=\proj{k},k\in[d]},\\
\label{cp2}\widehat{\sC}_{\cC}(\Phi):
&=&\max\Set{\cC(\Phi(\rho))-\cC(\rho):\rho \in \density{\cH}},
\end{eqnarray}
where $[d]$ denotes the set $\set{1,\ldots,d}$, and Equation \eqref{cp1} follows from the convexity of the measures of coherence \cite{Baumgratz2014}.

Finally, for a matrix $A\in\complex^{d\times d}$, we define its
matrix norm $\norm{A}_{1\rightarrow1}$ by \cite{Bhatia2013}
\begin{eqnarray*}
\norm{A}_{1\rightarrow1}:&=&\max\Set{\norm{Ax}_1,
\norm{x}_1=1}\\
&=&\max\Set{\sum^d_{i=1} \left|A_{ij}\right|:j=1,\ldots,d}.
\end{eqnarray*}
where $x=(x_1,x_2,\ldots,x_d)^T$ and $\norm{x}_{1}=\sum^{d}_{i=1}|x_{i}|$.

It turns out that
\begin{eqnarray}
\label{re:subprod}\norm{\prod_jA_j}_{1\to1}&\leqslant& \prod_j\norm{A_j}_{1\to1},\\
\label{re:eqten}\norm{\bigotimes_jA_j}_{1\to1}&=& \prod_j\norm{A_j}_{1\to1}.
\end{eqnarray}

\section{Operational interpretation of cohering power of quantum operations}
\label{sec:phy_mean}
By definition, the cohering power of a quantum operation $\Phi$  can be used to measure the maximal amount of coherence generated by $\Phi$. Besides,
based on an idea in the entanglement theory
\cite{Eisert06}, we consider the following operational task:
Given a quantum operation $\Phi:
\density{\cH}\rightarrow \density{\cH}$, if there exist states $\sigma$ and $\sigma'\in
\density{\cK}$ and an incoherent quantum operation $\Psi$ such that for any $\rho\in
\density{\cH}$,
\begin{eqnarray}
\Psi(\rho\ot \sigma)=\Phi(\rho)\ot \sigma',
\end{eqnarray}
then we say that the quantum operation $\Phi$ can be implemented by using an incoherent operation and an ancillary quantum system $\cK$.
[Note that here $\sigma$ and $\sigma'$ are not fixed. They can be different for different $\rho$.]
The question is, what is the minimal amount of coherence in the quantum state $\sigma$ ?
The following proposition answers this question, and provides a physical interpretation of cohering power of quantum operation $\Phi$.

\begin{prop}
Let $\Phi:\density{\cH}\rightarrow \density{\cH}$ be a quantum
operation and $\cC$ be a coherence measure. If $\Phi$ can be
implemented by using an incoherent operation $\Psi$ and an ancillary state $\sigma$ in quantum
system $\cK$, then the lower bound on the amount of coherence in the initial
state $\sigma$ of the ancillary system $\cK$ is $\sC_{\cC}(\Phi)$.
Moreover, if the coherence measure $\cC$ is subadditive, then the
lower bound on the amount of coherence in the state $\sigma$ is $\widehat
\sC_{\cC}(\Phi)$.
\end{prop}
\begin{proof}
Let $\sigma, \sigma'\in
\density{\cK}$ and $\Psi$ be an incoherent quantum operation such that for any $\rho\in
\density{\cH}$,
\begin{eqnarray*}
\Psi(\rho\ot \sigma)=\Phi(\rho)\ot \sigma'.
\end{eqnarray*}
Then, the coherence in $\rho\ot \sigma$ can not be less than $\Phi(\rho)\ot
\sigma'$. That is $\cC(\rho\ot \sigma)\geqslant \cC(\Phi(\rho)\ot
\sigma') $. Thus $\cC(\proj{k}\ot \sigma)\geqslant
\cC(\Phi(\proj{k})\ot \sigma')$ for any $k\in [d]$.
Since $\proj{k}$ is incoherent, the appending quantum operation with appending
state $\proj{k}$ is an incoherent operation. Moreover, the dismissal operation
is incoherent.
Therefore, $\cC(\sigma)\geqslant \cC(\proj{k}\ot \sigma)\geqslant
\cC(\Phi(\proj{k})\ot \sigma')\geqslant \cC(\Phi(\proj{k}))$. Thus
\begin{eqnarray*}\label{eq:lb1}
\cC(\sigma)\geqslant
\max\set{\cC(\Phi(\ket{k}\bra{k})):k\in[d]} = \sC_{\cC}(\Phi).
\end{eqnarray*}
Moreover, if the coherence measure $\cC$ is subadditive,
then
$\cC(\rho)+\cC(\sigma)\geqslant \cC(\rho\ot \sigma)\geqslant
\cC(\Phi(\rho)\ot \sigma')\geqslant \cC(\Phi(\rho))$. Hence
\begin{eqnarray*}\label{eq:lb2}
\cC(\sigma)\geqslant \max\set{\cC(\Phi(\rho)-\cC(\rho)):\rho \in
\density{\cH}} = \widehat \sC_{\cC}(\Phi),
\end{eqnarray*}
which implies that the lower bound on the amount of coherence in the state
$\sigma$ is $\widehat \sC_{\cC}(\Phi)$.
\end{proof}


\section{The calculation of cohering power of quantum operations}
\label{sec:cohe_pow}

In this section, we calculate  the cohering  power $\sC_{\cC}$ of
quantum operations, where $\cC$ denotes $\cC_{l_{1}}$ or $\cC_{r}$.
Among all quantum operations, the following three basic operations
are most important.

(i) \textbf{Unitary operation:}
Let $\Phi$ maps $\lin{\cH}$ to itself. If
there is a unitary operator $U$ such that for each $\rho\in D(\cH)$,
$$\Phi(\rho)=U\rho U^{\dag},$$ then $\Phi = \Phi_U$ is said to be a \emph{unitary operation}.

(ii) \textbf{Appending operation:}
Let $\Phi$ maps $\lin{\cH}$ to $\lin{\cH\ot\cK}$. If $\sigma\in
D(\cK)$ and for each $\rho\in D(\cH)$, $$\Phi(\rho)=\rho \ot \sigma,$$
then $\Phi = \Phi_A$ is said to be a \emph{appending operation}.

(iii) \textbf{Dismissal operation:}
Let $\{\cH_i\}_{i=1}^N$ be $N$ finite dimensional complex Hilbert
spaces, $\cH = \ot_{j=1}^N \cH_j$ is an $N$-partite system. If $\cH_0 =
\ot_{i=1}^{K_0} \cH_{j_i}$ with $K_0 \leq N$ and for each $\rho\in
\density{\cH}$, $$\Phi(\rho)=\mathrm{Tr}_{\cH_0}{\rho},$$ then $\Phi = \Phi_D$ is said to be
a \emph{dismissal operation}.

It is easy to show that the dismissal quantum operation
$\Phi_D$ is incoherent, and the appending quantum operation
$\Phi_A$ is incoherent if the appending state $\sigma$ is
incoherent.
Moreover, the following lemma shows that any quantum operation can be generated by the
above three quantum operations, that is

\begin{lem}[\cite{Dominique}]\label{prop:stine}
Let $\cH$ be a $d$-dimensional complex Hilbert space, and $\Phi:
\lin{\cH}\longrightarrow \lin{\cH}$ be a quantum operation.
Then there is a $d^2$-dimensional complex Hilbert space $\cK$, a pure state
$\ket{\psi} \in \cK$ and a unitary operator $U$ on $\cH\ot\cK$ such that for each $\rho\in \rD(\cH)$,
\begin{eqnarray}
\Phi(\rho)=\mathrm{Tr}_{\cK}\Pa{U(\rho \ot \proj{\psi})U^{\dag}}.
\end{eqnarray}
\end{lem}
Thus, the cohering power of above three basic quantum operations play an important role in
estimating the cohering power of a generic quantum operation.

\subsection{Cohering power of a quantum operation for $\cC_{l_{1}}$ measure}

Now, we compute the cohering power of above three basic quantum operations for $l_1$-norm measure of coherence, $\cC_{l_{1}}$.

\textbf{Unitary operation, $\Phi_U$:} Let $\Phi = \Phi_U$ be a unitary operation, where $U=(U_{ij})$ is a
unitary matrix. Since $\cC_{l_{1}}(U(\proj{k} U^{\dag}))
=\sum_{i\neq j}|(U\proj{k}U^{\dag})_{ij}|
=\sum^d_{i,j=1}|(U(\proj{k}U^{\dag})_{ij}|-1 = \Pa{\sum^{d}_{i=1}
|U_{ik}|}^{2}-1 $, then by the definitions of cohering power
$\sC_{\cC_{l_{1}}}$ and norm $\norm{\cdot}_{1\rightarrow
1}$,  we have
\begin{eqnarray}\label{CPU}
\sC_{\cC_{l_{1}}}(\Phi_U)=\norm{U}^2_{1\rightarrow1} - 1.
\end{eqnarray}

\textbf{Appending operation, $\Phi_A$:}
Let $\sigma\in \density{\cK}$, $\Phi_A(\rho)=\rho \otimes \sigma$. Then
\begin{eqnarray*}
\sC_{\cC_{l_{1}}}(\Phi_A)=\max\{ \cC_{l_{1}}(\delta \otimes \sigma):\delta \in
\cI\}=\cC_{l_{1}}(\sigma).
\end{eqnarray*}

\textbf{Dismissal operation, $\Phi_D$:}
Let $\Phi_D(\rho)=\mathrm{Tr}_{\cK}{\rho}$. Then
$$\sC_{\cC_{l_{1}}}(\Phi_D)=\max\Set{ \cC_{l_{1}}(\mathrm{Tr}_{\cK}{\delta}):\delta
\in \cI}=0.$$

By using \eqref{re:subprod}, \eqref{re:eqten} and (10), we can estimate the cohering power of global unitary operations
$\Phi_{\prod_jU_j}$ and $\Phi_{\bigotimes_j U_j}$ by each unitary operation $\Phi_{U_j}$, that is

\begin{prop}\label{prop:ten} Let $\{U_i\}$ be unitary operators on
$\cH$. Then

\end{prop}
\begin{eqnarray}\label{lUproduct}
\sC_{\cC_{l_{1}}}(\Phi_{\prod_j U_j})+1 \leqslant
\prod_j\Pa{\sC_{\cC_{l_{1}}}(\Phi_{U_j})+1},
\end{eqnarray}
\begin{equation}\label{lUtensor}
\sC_{\cC_{l_{1}}}(\Phi_{\bigotimes_j U_j})+1 =
\prod_j\Pa{\sC_{\cC_{l_{1}}}(\Phi_{U_j})+1}.
\end{equation}

Note that the property ~\eqref{lUtensor} was  obtained in
\cite{Mani2015}. Next, we show that the cohering power of unitary operations has continuous property, that is

\begin{prop} Let $\Phi_U$ and $\set{\Phi_{U_n}}_n$ be unitary operations on $\cH$. If $\lim_{n \rightarrow \infty}\norm{U_{n}-U}_{1\to1}=0$, then
\begin{equation*}
\lim_{n \rightarrow \infty} \sC_{\cC_{l_{1}}}(\Phi_{U_{n}}) = \sC_{\cC_{l_{1}}}(\Phi_U).
\end{equation*}
\end{prop}

\begin{proof} Let $\dim\cH=d$. Then for each unitary operator $V$ on $\cH$, $\norm{V}_{1\to 1}\leq \sqrt{d}$. Thus, the proposition follows from
\begin{eqnarray*}
&&|\sC_{\cC_{l_{1}}}(\Phi_{U_{n}})-\sC_{\cC_{l_{1}}}(\Phi_U)|\\
&=&| \norm{U_{n}}^{2}_{1\to1}-\norm{U}^{2}_{1\to1} |\\
  &\leqslant & | \norm{U_{n}}_{1\to1}-\norm{U}_{1\to1}|\cdot | \norm{U_{n}}_{1\to1}+\norm{U}_{1\to1} |\\
  &\leqslant &2\sqrt{d}| \norm{U_{n}}_{1\to1}-\norm{U}_{1\to1}|\\
  &\leqslant & 2\sqrt{d}\norm{U_{n}-U}_{1\to1}.
\end{eqnarray*} \end{proof}

In order to prove our main results in this section, we need the following proposition:

\begin{prop}\label{thm:impor}
Let $\set{\ket{k_i}: k_i\in[d_i]}_{i=1}^3$ be the reference bases of three complex Hilbert spaces $\{\cH_i\}_{i=1}^3$,
$\Phi_{1}:\lin{\cH_1} \longrightarrow \lin{\cH_2}$ and
$\Phi_{2}:\lin{\cH_2}\longrightarrow \lin{\cH_3}$ be quantum operations.
If $\Phi_2$ is a unitary operation, or an appending operation or a dismissal
operation, then
\begin{eqnarray}\label{cpproduct1}
\sC_{\cC_{l_{1}}}(\Phi_{2}\circ \Phi_{1})+1 \leqslant
(\sC_{\cC_{l_{1}}}(\Phi_{2})+1)\cdot(\sC_{\cC_{l_{1}}}(\Phi_{1})+1).
\end{eqnarray}
\end{prop}

\begin{proof} (i) If $\Phi_2=\Phi_{U}$ is a unitary operation, where $U=(U_{ij})$ is a unitary matrix, then
$\sC_{\cC_{l_1}}(\Phi_2)=\norm{U}^2_{1\rightarrow 1}-1$. Note that $\sum^{d_3}_{i,j=1}\abs{U_{ir}}|U_{js}|\leqslant \norm{U}^2_{1\rightarrow 1}=\sC_{\cC_{l_1}}(\Phi_2)+1$ and
\begin{eqnarray*}
&&\sC_{\cC_{l_{1}}}(\Phi_{2}\circ\Phi_{1})+1\\
&=&\cC_{l_{1}}\Pa{U(\Phi_1(\proj{k_1}))U^{\dag}}+1\\
&=&\sum^{d_3}_{i,j=1}\left|\sum^{d_2}_{r,s=1}U_{ir}(\Phi_1(\proj{k_1}))_{rs}U^{\dag}_{sj}\right|\\
&\leqslant&\sum^{d_3}_{i,j=1}\sum^{d_2}_{r,s=1}\left|U_{ir}\right|\left|U_{js}\right||\Phi_1(\proj{k_1})_{rs}|\\
&=&\sum^{d_2}_{r,s=1}\left|(\Phi_1(\proj{k_1}))_{rs}\right|\sum^{d_3}_{i,j=1}\abs{U_{ir}}|U_{js}|\\
&\leqslant& \Pa{\sC_{l_{1}}(\Phi_{2})+1}(\sC_{\cC_{l_{1}}}(\Phi_{1})+1).
\end{eqnarray*}
Hence, the inequality \eqref{cpproduct1} is proved.

(ii) If $\Phi_{2}$ is an appending operation, that is,
$\Phi_{2}: \rho \longrightarrow \rho \otimes \sigma$, then
$\sC_{\cC_{l_{1}}}(\Phi_{2})=\cC_{l_{1}}(\sigma) $, and
\begin{equation}
\cC_{l_{1}}(\Phi_{2}\circ
\Phi_{1}(\proj{k_{1}}))+1=\cC_{l_{1}}(\Phi_{1}(\proj{k_{1}}) \ot
\sigma)+1=(\cC_{l_{1}}(\Phi_{1}(\proj{k_{1}}))+1)\cdot(\cC_{l_{1}}(\sigma)+1).
\end{equation}
So we have $\sC_{\cC_{l_{1}}}(\Phi_{2} \circ\Phi_{1})+1 =
(\sC_{\cC_{l_{1}}}(\Phi_{2})+1)\cdot
(\sC_{\cC_{l_{1}}}(\Phi_{1})+1)$.\\

(iii) If $\Phi_{2}$ is a dismissal operation, then there is
a finite dimensional complex Hilbert space $\cK$ with a reference basis $\set{\ket{k_4}:
k_4\in[d_4]}$, such that
$\cH_{2}=\cH_{3}\ot \cK$ and
$\Phi_{2}(\rho)=\mathrm{Tr}_{\cK}{\rho}$. Note that
$\sC_{\cC_{l_{1}}}(\Phi_{2})=0$, and
\begin{equation}
\sC_{\cC_{l_{1}}}(\Phi_{1})=\max\Set{\sum^{d_4}_{r,u=1}\sum^{d_{3}}_{s,v=1}\abs{\Phi_{1}(\delta)_{rs,uv}}:
\delta=\left|k_1\rangle\langle k_1\right|, k_1\in[d_1]}.
\end{equation}
Hence, we have
\begin{eqnarray*}
&&\cC_{l_{1}}(\Phi_{2}\circ\Phi_{1}(|k\rangle\langle
k|))+1\\
&=&\sum^{d_{4}}_{r,u=1}|\sum^{d_{3}}_{s=1}\Phi_{1}(\ket{k_{1}}\bra{k_{1}})_{rs,us}|\\
&\leqslant&
\sum^{d_{4}}_{r,u=1}\sum^{d_{3}}_{s,v=1}|\Phi_{1}(\ket{k_{1}}\bra{k_{1}})_{rs,uv}|\\
&\leqslant& (\sC_{\cC_{l_{1}}}(\Phi_{1})+1)\cdot(\sC_{\cC_{l_{1}}}(\Phi_{2})+1).
\end{eqnarray*}
\end{proof}

Besides, due to  Lemma \ref{prop:stine}, Proposition \ref{thm:impor} and the zero cohering power of appending and dismissal operations, the cohering power of any quantum operation
is upper bounded by that of the corresponding unitary operation, where
the unitary operation is described in a larger dimensional space.
Another interesting proposition is the evaluation of
$\sC_{\cC_{l_{1}}}(\Phi_1\ot \Phi_2)$ in terms of $\sC_{\cC_{l_{1}}}(\Phi_1)$
and $\sC_{\cC_{l_{1}}}(\Phi_2)$, that is

\begin{prop}\label{prop:tencom}
Let $\cH_1, \cH'_1$, $\cH_2$ and $\cH'_2$ be finite dimensional
complex Hilbert spaces, $\set{\ket{k_1}: k_1\in[d_1]}$ and
$\set{\ket{k_2}: k_2\in[d_2]}$ be the reference basis of $\cH_1$ and
$\cH_2$, respectively, $\Phi_{1}:\lin{\cH_1}\longrightarrow
\lin{\cH'_1}$ and $\Phi_{2}:\lin{\cH_2}\longrightarrow \lin{\cH'_2}$
be two quantum operations. Then
\begin{eqnarray}\label{generaltensor}
\sC_{\cC_{l_{1}}}(\Phi_{1}\otimes
\Phi_{2})+1=(\sC_{\cC_{l_{1}}}(\Phi_{1})+1)\cdot(\sC_{\cC_{l_{1}}}(\Phi_{2})+1).~~~~~~~~~
\end{eqnarray}
\end{prop}

\begin{proof}
Note that for coherence measure $\cC_{l_1}$,
\begin{equation*}
\cC_{l_{1}}(\rho \ot \sigma)+1=
(\cC_{l_{1}}(\rho)+1)\cdot(\cC_{l_{1}}(\sigma)+1).
\end{equation*}
Therefore,
\begin{eqnarray*}
&&\sC_{\cC_{l_{1}}}(\Phi_1\ot \Phi_2)+1\\
&=&\max\{\cC_{l_1}(\Phi_{1}\ot\Phi_{2}(\delta)):\delta=\proj{k_1 k_2},
k_1\in[d_1], k_2\in[d_2]\}+1 \\
&=&\max\Set{\cC_{l_1}(\Phi_{1}(\delta_1))+1:\delta =\proj{k_1},k_1\in[d_1]}\times \\
&&\max\Set{\cC_{l_1}(\Phi_2(\delta_2))+1:\delta= \proj{k_2},k_2\in[d_2]}\\
&=&(\sC_{\cC_{l_{1}}}(\Phi_1)+1)\cdot(\sC_{\cC_{l_{1}}}(\Phi_2)+1).
\end{eqnarray*}
\end{proof}

\begin{remark}
In an $N$-qubit system, the cohering power of unitary operation
$\Phi_{H^{\ot N}}$, where
$H=\frac1{\sqrt{2}}\Br{\begin{array}{ccc}
1 & 1 \\
1& -1
\end{array}}$
is the Hadamard gate, is maximal over all unitary operations, that is,
$$
\sC_{\cC_{l_{1}}}\Pa{\Phi_{H^{\ot
N}}}=2^{N}-1=\max\Set{\sC_{\cC_{l_{1}}}(\Phi_U):U \in {\cal
U}\Pa{(\complex^2)^{\ot N}}}.
$$

We know that $H^{\ot N}$ plays an important role in quantum
computation and is the key to some quantum algorithms, such as
Shor's algorithm \cite{Shor94,Shor99}. Based on Solvay-Kitaev
theorem \cite{Kitaev2002}, any quantum algorithm can be realized
approximately by a sequence of unitaries from a finite set of
unitaries. This set is called a basis. For example, the set of gates
$$
Q=\Set{H,K,K^{-1},\Lambda(\sigma_{x}),\Lambda^{2}(\sigma_{x})},
$$
where
$K=\Br{\begin{array}{ccc}
1 & 0 \\
0 & i
\end{array}}$, $\Lambda(\sigma_{x})$ is the
controlled-not (CNOT) gate, and $\Lambda^{2}(\sigma_{x})$ is the CNOT
gate with two control qubits, constitutes a basis \cite{Kitaev2002}.
It is easy to see that the cohering
powers of all the gates in this basis is zero except the Hadamard gate.
So, the cohering power of a quantum algorithm can be determined
approximately by the number of Hadamard gates used in the algorithm.
\end{remark}

\subsection{Cohering power of a quantum operation for $\cC_{r}$ measure}

In this subsection, we will calculate the cohering power $\sC_{\cC_r}$ of
a generic quantum operation for relative entropy of coherence. Firstly, we compute also the cohering power of three basic quantum operations for  $\cC_{r}$.\\

(i) For unitary operation $\Phi_U$, we consider the following two
special cases to witness some features of $\sC_{\cC_{r}}(\Phi_U)$.

$\bullet$ $U$ on single-qubit system, that is, $U$ is a $2\times 2$
unitary matrix. Since $U$ can be represented  by
\begin{equation*}\label{7}
U= e^{\mathrm{i}\varphi}\Br{\begin{array}{ccc}
a & b  \\
-b^{\ast} & a^{\ast}
\end{array}},
\end{equation*}
where $|a|^{2}+|b|^{2}=1$, we have
\begin{equation*}
\sC_{\cC_{r}}(\Phi_U)=-|a|^{2}\log_{2}|a|^{2}-|b|^{2}\log_{2}|b|^{2}= \rS\Pa{
|a|^{2},|b|^{2}}.
\end{equation*}

$\bullet$ $U_{AB} = U_{A}\otimes U_{B}$ on two-qubit system,
where $U_A$ and $U_B$ are $2\times 2$ unitary matrices, we have
\begin{equation*}
\sC_{\cC_{r}}(\Phi_{U_{AB}})=\rS(\rho_{\diag})=\rS\Pa{
|a|^{2},
|b|^{2}
}+\rS\Pa{
|c|^{2}, |d|^{2}}.
\end{equation*}

After some simple calculation, one can obtain the cohering power
$\sC_{\cC_{r}}$ of a general unitary operation $\Phi_U$ with
$U=(U_{ij})$, as
\begin{eqnarray}
\sC_{\cC_{r}}(\Phi_{U})=\max\set{S(|U_{1i}|^2, |U_{2i}|^2,
\cdots, |U_{di}|^2),i\in[d]}.~~~~~~~~
\end{eqnarray}

(ii) For appending operation $\Phi_A$, that is, $\Phi_A: \rho
\longrightarrow \rho \ot \sigma$, we have
$\sC_{\cC_{r}}(\Phi_A)=\max \Set{\cC_{r}(\Phi_A(\delta)):\delta \in
\cI}=\max\Set{\cC_{r}(\delta \ot \sigma):\delta \in
\cI}=\cC_{r}(\sigma)$.

(iii) For dismissal operation $\Phi_D$, that is $\Phi: \rho
\longrightarrow \mathrm{Tr}_{\cK}{\rho}$, we have
$\sC_{\cC_{r}}(\Phi_D)=\max\Set{\cC_{r}(\mathrm{Tr}_{\cK}{\delta}):\delta \in
\cI}=0$.\\

Though the cohering power of three basic quantum operations have
been obtained for relative entropy of coherence, it is difficult to use the same prescription
as in $\cC_{l_1}$
measure case to deal with $\cC_r$ measure case. However, by the
definition of $\sC_{\cC_r}$, we still have the following proposition
which allows us to evaluate the cohering power of $\ot_j\Phi$ in terms of that
of each $\Phi_j$.

\begin{prop}
Let $\{\Phi_j\}^n_{j=1}$ be a series of quantum operations. Then
\begin{eqnarray}\label{addre}
\sC_{\cC_{r}}\Pa{\bigotimes^n_{j=1} \Phi_j} = \sum^n_{j=1}\sC_{\cC_{r}}(\Phi_j).
\end{eqnarray}
\end{prop}

\section{Comparison between $\sC_{\cC}$ and $\widehat\sC_{\cC}$}
\label{sec:diff_def}

In this section, we compare the cohering power $\sC_{\cC}$ and
$\widehat\sC_{\cC}$ of quantum operations. By the definitions,
$\sC_{\cC}(\Phi)$ is always less than $ \widehat\sC_{\cC}(\Phi)$,
where $\cC$ denotes the coherence measure $\cC_{l_1}$ or $\cC_{r}$.
For simplicity, we consider only unitary operations. Firstly, we
study the $\cC_{l_1}$ measure case and find that $\sC_{\cC}(\Phi)$ and $ \widehat\sC_{\cC}(\Phi)$ can be equal in single qubit systems.

\begin{prop}\label{prop:com1}
The cohering powers $\sC_{\cC_{l_1}}$ and
$\widehat\sC_{\cC_{l_1}}$, of any unitary operation $\Phi_U$ on a single qubit system, are equal.
However, for an N-qubit system ($N \geqslant 2$), there exists a unitary operation $\Phi_{U_N}$
such that $\sC_{\cC_{l_1}}(\Phi_{U_N})<\widehat\sC_{\cC_{l_1}}(\Phi_{U_N})$.
\end{prop}

\begin{proof}
(1). In single-qubit case, any unitary operator $U$ can be written as $
U= e^{i \varphi}\Br{\begin{array}{ccc}
a & b  \\
-b^{\ast} & a^{\ast}
\end{array}}
$ with $|a|^{2}+|b|^{2}=1$, and any state $\rho\in D(\complex^2)$
can be expressed as $\rho=\frac{\mathbb{I}}{2}+\frac{1}{2}
\vec{r}\cdot{\vec{\sigma}}$, where $\vec{r}=(x, y, z)$ is a unit
vector and ${\vec{\sigma}}=(\sigma_x, \sigma_y, \sigma_z)$ are the
Pauli matrices. Thus,
\begin{eqnarray*}
\sC_{\cC_{l_1}}(\Phi_U)=\max\Set{\cC_{l_{1}}(\Phi_U(\delta)):\delta\in \cI}=2\abs{ab},
\end{eqnarray*}
and
\begin{eqnarray*}
&&\cC_{l_{1}}(U\rho U^{\dag})-\cC_{l_{1}}(\rho)\\
&=&\abs{a^{2}(x-\mathrm{i}y)+b^2(x+\mathrm{i}y)-2abx} - \abs{x - \mathrm{i}y}\\
&\leqslant&|a|^{2}\abs{x-\mathrm{i}y} + |b|^{2}\abs{x+\mathrm{i}y} + 2|ab||x| - \abs{x-\mathrm{i}y}\\
&\leqslant&2\abs{ab}\abs{x}\leqslant\sC_{\cC_{l_1}}(\Phi_U),
\end{eqnarray*}
where the last inequality is due to $|x|\leqslant 1$. Hence we have $\widehat\sC_{\cC_{l_1}}(\Phi_U)\leqslant\sC_{\cC_{l_1}}(\Phi_U)$. However, as $\widehat\sC_{\cC_{l_1}}(\Phi_U)$ is always larger than  $\sC_{\cC_{l_1}}(\Phi_U)$, we have $\sC_{\cC_{l_1}}(\Phi_U)=\widehat\sC_{\cC_{l_1}}(\Phi_U)$ for any
unitary operation $\Phi_U$.

(2). For any $U\in {\cal U}(\complex^2)$, it follows from (1) that
$\sC_{\cC_{l_1}}(\Phi_U)=\widehat\sC_{\cC_{l_1}}(\Phi_U)$.
Thus there exists $ \rho_{1} \in \density{\complex^2}$ such
that $\sC_{\cC_{l_1}}(\Phi_U)=\cC_{l_{1}}(U\rho_{1}
U^{\dag})-\cC_{l_{1}}(\rho_{1})$. Now, we take a quantum state
$\sigma \in \density{\cH_{N-1}}$ such that $\cC_{l_{1}}(\sigma)>0$ and $\rho_1 \ot \sigma \in \density{\cH_N}$. Let
$U_{N}=U_1 \ot \mathbb{I}_{N-1}$ and $\rho_{N}=\rho_{1}\ot \sigma$, where
$\mathbb{I}_{N-1}$ denotes the identity operator on the remaining ($N-1$)-qubit
system. Then
\begin{equation*}
\sC_{\cC_{l_1}}(\Phi_{U_{N}})=\sC_{\cC_{l_1}}(\Phi_{U_1}),
 \end{equation*}
and
\begin{eqnarray*}
&& \cC_{l_{1}}(U_{N}\rho_{N} U^{\dag}_{N})-\cC_{l_{1}}(\rho_{N})\\
&=&\cC_{l_{1}}(U_{1}\rho_{1}U^{\dag}_{1}\otimes \sigma)-\cC_{l_{1}}(\rho_{1}\otimes \sigma)\\
&=&(\cC_{l_{1}}(U_{1}\rho_{1}U^{\dag}_{1})+1)\cdot(\cC_{l_{1}}(\sigma)+1)
-(\cC_{l_{1}}(\rho_{1})+1)\cdot(\cC_{l_{1}}\sigma)+1)\\
&=&(\cC_{l_{1}}(U_{1}\rho_{1} U^{\dag}_{1})-\cC_{l_{1}}(\rho_{1}))\cdot(\cC_{l_{1}}(\sigma)+1)\\
&>&\cC_{l_{1}}(U_{1}\rho_{1} U^{\dag}_{1})-\cC_{l_{1}}(\rho_{1})\\
&=&\sC_{\cC_{l_1}}(\Phi_{U_{1}}).
\end{eqnarray*}
Therefore,
\begin{equation*}
\sC_{\cC_{l_1}}(\Phi_{U_{N}})< \widehat
\sC_{\cC_{l_1}}(\Phi_{U_{N}}).
\end{equation*}
\end{proof}

It follows from Proposition \ref{prop:com1} that
$\sC_{\cC_{l_1}}(\Phi_U)=\widehat\sC_{\cC_{l_1}}(\Phi_U)$ for any
unitary operation $\Phi_U$ on a single qubit system. This implies that the
maximal coherence, produced by $\Phi_U$ over all states, can be obtained
by considering only basis states. Moreover, the following proposition
shows that for any unitary operation on any quantum system,
$\sC_{\cC_{l_1}}(\Phi_U)$ can describe the maximal coherence
produced over all states in some sense.

\begin{prop}
The cohering power $\sC_{\cC_{l_1}}$ of any unitary operation $\Phi_U$, on a $d$-dimensional quantum system $\cH_d$, is given by
\begin{eqnarray}
\sC_{\cC_{l_1}}(\Phi_U)=\max \Set{\frac{\cC_{l_{1}}(U \rho
U^{\dag})-\cC_{l_{1}}(\rho)}{\cC_{l_{1}}(\rho)+1}:\rho\in
\density{\cH_d}}.~~~~~~~~~~~~~~~~
\end{eqnarray}
\end{prop}

\begin{proof}
Since
\begin{eqnarray*}
&&\sum^d_{i,j=1}\abs{(U \rho
U^{\dag})_{ij}}=\sum^d_{i,j=1}\abs{\sum^d_{s,t=1}U_{is} \rho_{st}
U^{\dag}_{tj}}\\
&\leqslant& \sum^d_{i,j=1}
\sum^d_{s,t=1}\abs{U_{is}}\abs{\rho_{st}}\abs{U_{jt}}=
\sum^d_{s,t=1}\abs{\rho_{st}}\Pa{\sum^d_{i,j=1}\abs{U_{is}}\abs{U_{jt}}}\\
&\leqslant&\Pa{\norm{U}^{2}_{1\to1}}\sum^d_{s,t=1}\abs{\rho_{st}},
\end{eqnarray*}
therefore
\begin{eqnarray*}
\cC_{l_{1}}(U \rho U^{\dag})-\cC_{l_{1}}(\rho)
&\leqslant&(\norm{U}^{2}_{1\to1}-1)\sum^d_{i,j=1}\abs{\rho_{ij}}
=\sC_{\cC_{l_{1}}}(\Phi_U)\cdot(\cC_{l_{1}}(\rho)+1).
\end{eqnarray*}
As the above expression is true for any $\rho$, we have
\begin{eqnarray*}
&&\sC_{l_{1}}(\Phi_U)\geqslant \max\Set{\frac{\cC_{l_{1}}(U \rho
U^{\dag})-\cC_{l_{1}}(\rho)}{\cC_{l_{1}}(\rho)+1}:\rho\in
\density{\cH_d}}.
\end{eqnarray*}
On the other hand, it is clear that
\begin{eqnarray*}
\sC_{l_{1}}(\Phi_U)\leqslant\max\Set{\frac{\cC_{l_{1}}(U \rho
U^{\dag})-\cC_{l_{1}}(\rho)}{\cC_{l_{1}}(\rho)+1}:\rho\in
\density{\cH_d}}.
\end{eqnarray*}
Hence the proof is completed.
\end{proof}

The above proposition implies that $\sC_{\cC_{l_1}}(\Phi_U)\cdot(\cC_{l_{1}}(\rho)+1)\geq\cC_{l_{1}}(U \rho U^{\dag})-\cC_{l_{1}}(\rho)$ for
any quantum state $\rho$. That is, given a unitary operation $\Phi_U$ and a quantum state $\rho$, the
coherence produced after the operation of $\Phi_U$ on $\rho$, is determined by the coherence of the initial state $\rho$ and the
cohering power  $\sC_{\cC_{l_1}}(\Phi_U)$, although the cohering power $\sC_{\cC_{l_1}}$ is defined only for the
incoherent states (see Eq.~\eqref{cp3}). Next, we
compare  $\sC_{\cC_{r}}(\Phi_U)$ with $\widehat\sC_{\cC_{r}}(\Phi_U)$. Note that, in general, the features of cohering powers for $l_1$ norm of coherence may not hold true for relative entropy of coherence.

\begin{prop}\label{prop:compr}
For an $N$-qubit system
($N\geqslant 1$), the cohering powers $\sC_{\cC_{r}}$
and $\widehat\sC_{\cC_{r}}$ of any unitary operation $\Phi_U$ are not equal, in general. That is,
there exists a unitary operation $\Phi_{U_N}$ such that
$\sC_{\cC_{r}}(\Phi_{U_N})<\widehat\sC_{\cC_{r}}(\Phi_{U_N})$.
\end{prop}

\begin{proof}
First, we give a specific example which shows that $\sC_{\cC_{r}}(\Phi_{U_N})$ and $\widehat\sC_{\cC_{r}}(\Phi_{U_N})$ are not equal even for a single qubit system. Let
\begin{equation*}\label{counterU}
U_{1}=
\left( \begin{array}{ccc}
  0.5828-0.8125i & -0.0148+0.0007i\\
  -0.0125-0.0080i & -0.1021-0.9947i
\end{array} \right),
\end{equation*}
and
\begin{equation*}\label{7}
\rho_{1}=
\left( \begin{array}{ccc}
   0.8706    &   0.3078+0.0527i\\
   0.3078-0.0527i &  0.1294
\end{array} \right).
\end{equation*}

Then $\sC_{\cC_{r}}(\Phi_{U_{1}})\approx 0.0030$ and
$[\cC_{r}(U_{1}\rho_{1} U^{\dag}_{1})-\cC_{r}(\rho_{1})]\approx
0.0190$. Thus, $\sC_{\cC_{r}}(\Phi_{U_{1}}) < \widehat\sC_{\cC_{r}}(\Phi_{U_1})$, for a single qubit system.
[Note that $U_1$ and $\rho_1$ above are chosen randomly using a computer program. Hence, they have this complex numerical form. However, it could be possible to construct a simple unitary to demonstrate the same.]

In $N$-qubit case, we take $U_{N}=U_{1}\otimes \mathbb{I}_{N-1}$ and
$\rho_{N}=\rho_{1}\otimes \sigma$, where $\sigma$ is a state of the
($N-1$)-qubit system. Then
 \begin{equation*}
 \sC_{\cC_{r}}(\Phi_{U_{N}})=\sC_{\cC_{r}}(\Phi_{U_{1}}),
 \end{equation*}
and
\begin{eqnarray*}
 &&\cC_{r}(U_{N}\rho_{N} U^{\dag}_{N})-\cC_{r}(\rho_{N})\\
&=&\cC_{r}(U_{1}\rho_{1}U^{\dag}_{1}\otimes \sigma)-\cC_{r}(\rho_{1}\otimes \sigma)\\
&=&\cC_{r}(U_{1}\rho_{1}U^{\dag}_{1})+\cC_{r}(\sigma)
 -(\cC_{r}(\rho_{1})+\cC_{r}(\sigma))\\
&=&\cC_{r}(U_{1}\rho_{1} U^{\dag}_{1})-\cC_{r}(\rho_{1}).
\end{eqnarray*}

Since
$\sC_{\cC_{r}}(\Phi_{U_{N}})=\sC_{\cC_{r}}(\Phi_{U_{1}})<\cC_{r}(U_{1}\rho_{1} U^{\dag}_{1})-\cC_{r}(\rho_{1})=\cC_{r}(U_{N}\rho_{N} U^{\dag}_{N})-\cC_{r}(\rho_{N})$, we have
\begin{equation*}
\sC_{\cC_{r}}(\Phi_{U_{N}})<\widehat\sC_{\cC_{r}}(\Phi_{U_{N}}).
\end{equation*}
\end{proof}

Proposition \ref{prop:compr} shows that above two cohering
powers defined for $\cC_r$ measure are not equal even in a single qubit
system, which is different from the $l_1$-norm case. Thus, we need to
specify the coherence measure when we refer to the cohering power.

\section{Conclusion}
\label{sec:conclu}
A major bottleneck of quantum technology is the quantum decoherence effect, which leads to the dissipation from quantum systems to classical systems. Quantum correlations, especially entanglement, are fragile under decoherence. However, due to vast applications of quantum correlations and entanglement in quantum information processing tasks, maintaining coherence and/or quantum correlations in quantum systems is a desirable task. We must check the phenomenon of decoherence when we implement quantum techniques in quantum information and computation. Here, the role of cohering power of quantum operations becomes crucial. 
In this work, we have investigated the cohering power of
generic quantum operations and compared two different types of cohering power. First,
 we provided an information-theoretic interpretation of cohering power of quantum operations. We showed
that the minimal amount of coherence of an ancillary quantum state such
that a given quantum operation can be implemented by using an incoherent
operation and an ancillary quantum system is just its cohering power.
Moreover, by dividing a generic quantum operation into three basic quantum operations, 
namely, unitary operation, appending operation and dismissal operation, we showed that 
the cohering power of any quantum operation is upper bounded by that of the corresponding unitary operation.
Furthermore, we compared
two different kinds of cohering powers ($\sC_{\cC}$ and
$\widehat{\sC}_{\cC}$) for unitary operations. For $\cC_{l_1}$
measure, they coincide in single qubit case only, while they are
different for any number of qubits in the case of $\cC_r$ measure.

\smallskip
\noindent
\begin{acknowledgments}
AK acknowledges the research fellowship of Department of Atomic Energy, Government of India.
K. Bu thanks Uttam Singh and Akihito Soeda for their useful
comments and suggestions. L. Zhang is supported by the National Natural Science
Foundation of China (No.11301124), and J. Wu is supported
by National Natural Science Foundation of China (No. 11171301,
11571307) and by the Doctoral Programs Foundation of the Ministry of
Education of China (J20130061).
We also thank anonymous Referees for their insightful comments, which resulted in important changes in the manuscript.
\end{acknowledgments}

\end{document}